\documentclass[12pt]{article}
\usepackage[margin=0.7in]{geometry}
\usepackage{float}
\usepackage{amsmath}
\usepackage{amssymb}
\usepackage{graphicx}
\usepackage{cite}
\usepackage{color}
\usepackage{dcolumn}
\usepackage{bm}
\usepackage{newfloat}
\DeclareFloatingEnvironment[name={SI Figure}]{suppfigure}
\DeclareFloatingEnvironment[name={SI equation}]{suppequation}
\DeclareFloatingEnvironment[name={SI Table}]{supptable}

\usepackage{amsmath}
\usepackage{graphicx}
\usepackage{dcolumn}
\usepackage{bm}
\usepackage{amsmath}
\usepackage{amssymb}
\usepackage{amsfonts}
\usepackage{graphicx}
\usepackage{dcolumn}
\usepackage{bm}
\usepackage{sidecap}
\usepackage {caption}
\usepackage{float}
\usepackage{parskip}

\usepackage{hyperref}
\usepackage{longtable}
\usepackage{array}
\usepackage{array,booktabs,arydshln,xcolor}

\usepackage{booktabs}

\usepackage{graphics}
\usepackage{xcolor,colortbl}

\definecolor{Gray}{gray}{0.2}
\definecolor{LightCyan}{rgb}{0.2,0.2,1}

\newcolumntype{a}{>{\columncolor{Gray}}c}
\newcolumntype{b}{>{\columncolor{white}}c}

\begin{document}

\begin{flushleft}
{\large
\textbf{A multilayer PPI network analysis of different life stages in C. elegans}
}
\\
Pramod Shinde$^{1}$, and Sarika Jalan$^{1,2,*}$
\\ 
\it ${^1}$ Centre for Biosciences and Biomedical Engineering, Indian Institute of Technology Indore, Simrol, Indore 452020, India\\
\it ${^2}$ Complex Systems Lab, Discipline of Physics, Indian Institute of Technology Indore, Simrol, Indore 452020, India\\

E-mail: pramodshinde119@gmail.com, $^{\ast}$sarikajalan9@gmail.com 
\end{flushleft}

\begin{abstract}
Molecular networks act as the backbone of cellular activities, providing an {excellent} opportunity to understand the developmental changes in an organism.
While network data usually constitute only stationary network graphs, constructing multilayer PPI network may provide clues to the particular developmental role at each {stage of life} and may unravel the importance of these developmental changes.
The developmental biology model of {\it Caenorhabditis elegans} {analyzed} here provides a ripe platform to understand the patterns of evolution during life stages of an organism.
In the present study, the widely studied network properties exhibit overall similar statistics for all the PPI layers.
Further, the analysis of the degree-degree correlation and spectral properties not only reveals crucial differences in each PPI layer but also indicates the presence of the varying complexity among them.
The PPI layer of Nematode life stage exhibits various network properties different to rest of the PPI layers, indicating the specific role of cellular diversity and developmental transitions at this stage.
The framework presented here provides a direction to explore and understand developmental changes occurring in different life stages of an organism.
\end{abstract}

Recent developments in the quantitative analysis of complex networks have rapidly been translated to studies of different biological network organizations \cite{Lazer}.
Developmental biology is the study of the molecular and cellular events that lead to the generation of a multicellular organism from a fertilized egg \cite{West_Eberhard}. 
Although much is known about the morphological changes that take place during the development, there is a lesser understanding of the mechanisms by which these changes occur.
Due to lack of this knowledge, and because of the interest in understanding how something as complex as a living organism can develop from a single cell, developmental biology is one of the most active areas of biological research today \cite{Raff}.
The intimidating complexity of cellular systems appears {to be} a major hurdle in understanding internal organization of molecular pathways and their development in large scale evolutionary biological networks \cite{Schuster}.
For instance, during the development phase of {\it Caenorhabditis elegans (C. elegans)} from undeveloped embryo to completely developed nematode, it undergoes multiple physiological and physiochemical changes.
Many key discoveries, both in basic biology and medically relevant areas, were first made in the worm. 
Since its introduction, {\it C. elegans} has been used to study a much larger variety of biological processes \cite{Brenner}.
Together, these studies revealed a surprisingly strong conservation in molecular and cellular pathways between worms and mammals. 
Indeed, subsequent comparison of the human and {\it C. elegans} genomes confirmed that the majority of human disease genes and disease pathways are present in {\it C. elegans} \cite{Kuwabara}.
Here, the global architecture of protein-protein interaction (PPI) network of each life stage of this model system is considered and focus is given to understand the behavior of individual layer of this multilayer network.

It is worth noticing that layers of a multilayer network can be tackled from different perspectives \cite{rev_multi_layer} and might in principle be used to understand the developmental biological changes in different life stages of this organism.
Recently, following information from theoretical and statistical mechanics paradigms, several {structural and spectral} measures for randomness and complexity have been proposed for social, technological and biological {networks} \cite{Ferrer, Dorogovtsev1, Vattay, Clauset}.
These measures have been shown to be extremely successful in quantifying the level of organization encoded in structural features of networks.
The measures like degree-degree correlation and von Neumann entropy allow us to capture differences and similarities between networks, which furthers our understanding of the information encoded in complex networks \cite{Anand}.
This complexity resides not only in the sheer number of proteins and interactions taking part in particular layer of multilayer network, but also in how individual layer is evolved or designed to fulfill the cellular functions.
{To understand this,} an extend of varying randomness and complexity is deduced using degree-degree correlation and spectral properties of each layer of the multilayer PPI network.
Further, the early Nematode developmental stage is more complex among all the life stages of {\it C. elegans}.
This analysis provides a direction to understand and capture important developmental changes in an organism.

\section*{Methods}
\subsection*{Network construction}

\begin{figure}[h]
\begin{center}
\includegraphics[width=12cm,height=7cm]{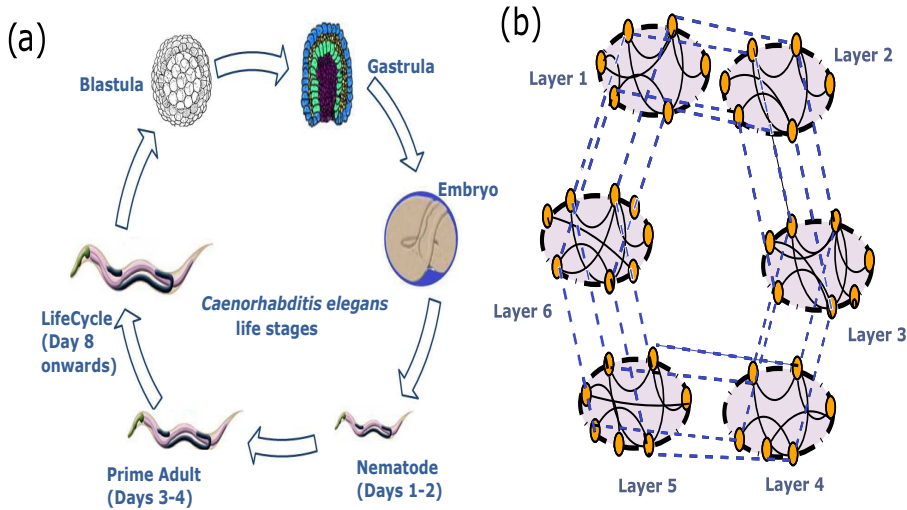}
\caption{(Color On-line) {Schematic representation of {\it C. elegans} {life cycle} and multilayer network architecture.} (a) Six different life stages of {\it C. elegans}. (b) Multilayer network architecture where nodes (circles) have connections (rigid lines) between them in a layer. Note that there may be different set of nodes appearing in different layers and dotted lines represent common nodes present across all the layers. }
\label{fig1}
\end{center}
\end{figure}

After extracting the names of proteins occurring in a particular life stage of {\it Caenorhabditis elegans}, experimentally verified protein-protein interactions { are collected} from various bioinformatics  repositories \cite{SM}. 
Next, six different PPI networks {are constructed} by treating each life stage as a layer of the multilayer PPI network (Fig. \ref{fig1}). 
For each layer, all the proteins {are enlisted and these proteins are} occurring for their functional or structural activities in that life stage.
The proteins are the nodes and connections are assigned if a pair of proteins $i$ and $j$ has an interaction between them.
Thus, six different sub-networks for multilayer PPI network {are obtained}.
The adjacency matrix of each layer of the multilayer network is denoted as $A_\alpha$ and elements are defined as,

\begin{equation}
A_{\mathrm {ij}}^\alpha = \begin{cases} 1 ~~\mbox{if } i \sim j \\
0 ~~ \mbox{otherwise} \end{cases}
\label{adj_wei}
\end{equation}

All the adjacency matrices are symmetric ({\it i.e.,} $A_{ij}^{\alpha}$ = $A_{ji}^{\alpha}$) where $\alpha = 1,2,...,6$.

\subsection*{Structural properties}

The most basic structural parameter of a network is the degree of a node ($k_i$), which is defined as a number of edges connected to the node ($k_i=\sum_j A_{ij}$).
The degree distribution, $p(k)$, {is calculated} which is the probability that a randomly chosen node has $k$ connections.
The second parameter, the clustering coefficient $(C)$, is the ratio of the number of interactions a neighbor of particular node is having and the possible number of connections the neighbors can have among themselves.
Further, the network diameter ($D$) is defined as the longest of the shortest paths between all the pair of nodes in a network \cite{Watts}.
Another property of the network which turns out to be crucial in distinguishing the individual layer of the multilayer PPI network is the Pearson degree-degree correlation ($r$), which measures the tendency of nodes with the similar numbers of edges to connect.
It can be defined as \cite{Rivera,Disso},
 
 \begin{equation}
r = \frac{[\frac{1}{M} \sum_{i} j_i k_i] - [ \frac{1}{M} \sum_i \frac{1}{2}(j_i + k_i)^2]}
{[ \frac{1}{M} \sum_{i} \frac{1}{2} (j_i^2+ k_i^2)] - [ \frac{1}{M} \sum_i \frac{1}{2}(j_i + k_i)^2]}
\label{assortativity}
 \end{equation}
 
where $j_{i}$ and $k_{i}$ are the degrees of the nodes connected through the $i^{th}$ edge, and $M$ is the number of edges in the network.
The value of $r$ being zero corresponds to a random network where as the negative(positive) values correspond to dis(assortative) networks.

Further, correlation between link betweenness centrality and overlap of the neighborhood of two connected nodes, {is calculated}.
Link betweenness centrality ($\beta_{L}$) is defined for an undirected link as,
$ \beta_{L} = \sum_{v \in V_{s}} \sum_{w \in V/{v}} \sigma_{vw} (e)/\sigma_{vw}$
 where $\sigma_{vw} (e)$ is the number of shortest paths between $v$ and $w$ that contain $e$, and $\sigma_{vw}$ is the total number of shortest paths between $v$
and $w$. 
The overlap of the neighborhood ($O_{ij}$) of two connected nodes $i$ and $j$ is defined as,
$O_{ij} = \frac{n_{ij}} {(k_{i} - 1) + (k_{j} - 1) - n_{ij}}$
where $n_{ij}$ is the number of neighbors common to both nodes $i$ and $j$ \cite{Onnela}. Here $k_i$ and $k_j$ represent the degree of the $i^{th}$ and $j^{th}$ nodes .
Further, Pearson correlation coefficient ($O\beta_L$) of $O_{ij}$ and $\beta_{L}$ {can be defined} as,
\begin{equation}
O\beta_L= \frac{( O_{i,j}- \textless O_{i,j} \textgreater) (\beta_L - \textless \beta_L \textgreater)}{\sqrt{( O_{i,j}- \textless O_{i,j} \textgreater)^2} \sqrt{(\beta_L - \textless \beta_L \textgreater)^2}}
\label{lap_eigval_entp}
\end{equation}

In particular, negative value of $O\beta_L$ coefficient suggests the importance of weak ties, a concept borrowed from social sciences into the network analysis.

\subsection*{Spectral properties}
The eigenvalues of the adjacency matrix {are denoted} by $\lambda_i$, i = $1, 2,  . . . , N$ such that λ$\lambda_1 < \lambda_2 < \lambda_3 < . . . < \lambda_N$.
The duplicated nodes in a network can be identified from corresponding adjacency matrix in the following manner. 
When (i) two rows (columns) have exactly same entries, it is termed as the complete row (column) duplication {\it i.e.} $R1 = R2$, (ii) when a combination of rows (columns) have exactly same entries as another combination of rows (columns) then it is termed as the partial duplication of rows (columns), for example $R1 = R2+R3$. 
Satisfying any one of the conditions (i) and (ii) lowers the rank of the matrix exactly by one. 
In addition, the rank is also lowered if (iii) there is an isolated node.
All these conditions lead to the zero eigenvalues in the matrix spectra \cite{BookSpectra,Alok}.
Since there is no isolated node in the network, ensured in the beginning itself by considering only the largest connected cluster for this analysis, conditions (i) and (ii) are the only conditions responsible for occurrence of {the} zero degeneracy.

Further, the von Neumann entropy $S_{\alpha}$ of the graph $G$ {is calculated } as, $S_{\alpha} = - Tr [\mathcal{L}_{G} \hspace{0.1cm} log_2 \hspace{0.1cm} \mathcal{L}_{G}]$ where $\mathcal{L}_G = c \times (D-A)$ is the combinatorial Laplacian matrix \cite{Laplacian} and D is the diagonal matrix of the degrees re-scaled by $c=\frac{1}{\sum_{i, j \epsilon V} a_{i,j}} = \frac{1}{2N_C}$.
Formally, $\mathcal{L}_G$ has all the properties of a density matrix {\it i.e.,} it is  positive semi-definite and $Tr (\mathcal{L}_G) = 1$ \cite{Domenico}. 
Therefore, $S_{\alpha}$ can be written as,

\begin{equation}
S_{\alpha}= - \sum_{i=1}^{N} \lambda_i^L \log_2 (\lambda_i^L)
\label{lap_eigval_entp}\end{equation}

The maximum entropy, a network of size $N$ can achieve, is $log N$.

Further, {the network} properties {are compared between} PPI layers {and} corresponding Erd\H os R\' enyi (ER) random  networks  \cite{BarabasiReview}. 
This allows to estimate the probability that a random network with certain constraints has of belonging to a particular architecture, and thus assess the relative importance of different network architecture and help discern the mechanisms responsible for given real-world networks \cite{Johnson}.

\section*{Results and discussions}
\subsection*{Structural properties}

\begin{table}
\caption{{The properties of different layers of multilayer PPI network of {\it C. elegans} life stages.} 
Here $N$, $N_{C}$, $D$, $\langle K \rangle$, $\langle C \rangle$, $r$, $\lambda_{0}$, $D_{0}$, $O\beta_L$ and $S$ represent number of participating nodes, number of connections, diameter, average degree, average clustering coefficient, degree-degree correlation coefficient, number of zero eigenvalues, number of complete duplicates, overlap-link betweeness correlation coefficient, and von Neumann entropy coefficient, respectively.}
\begin{tabular} {c c c c c c c c c c c}   
\hline  \specialrule{1pt}{0pt}{0pt}
Layer          & $N$       & $N_{C}$  &$D$ & $\langle K \rangle$   & $\langle C \rangle$    & $r$          &$\lambda_{0}(\%)$ &$D_{0}(\%)$&$O \beta_L$ &$S$\\ \hline  \specialrule{1pt}{0pt}{0pt}
Blastula 	    & 2876   	& 22880    &12  & 16	 	                & 0.24                 & 0.24   &812(28.2)         &716(24.9)		   &-0.39      &9.68\\ 
Gastrula   	& 2848     	& 22802    &12  & 16	 	                & 0.25                 & 0.24   &791(27.7)         &692(24.3)		   &-0.39      &9.73\\  
Embryo   	& 3568    	& 25741    &12  & 14	 	                & 0.23                 & 0.27   &1144(32.1)       &1025(28.7)		&-0.37     &9.18\\
Nematode 	& 4755     	& 35708    &12  & 15	 	                & 0.33                 & 0.16   &2087(43.9)       &1928(40.5)		&-0.30     &9.99\\
PrimeAdult	& 3112     	& 24126    &12  & 16	 	                & 0.24                 & 0.26   &926(29.8)         &833(26.8)		    &-0.38     &9.59\\ 
LifeCycle 	& 3415     	& 25255    &12  & 15	 	                & 0.23                 & 0.27   &1057(30.9)       &937(27.4)		    & -0.36    &9.33\\ \hline \specialrule{1pt}{0pt}{0pt}
\end{tabular}
\begin{flushleft} 
\label{table1} 
\end{flushleft}
\end{table}

First, the structural properties of PPI networks {are analyzed} for six developmental stages. 
The average degree, which gives a measure of the average connectivity of individual network, remains same for all the layers of the multilayer PPI network (Table \ref{table1}).
This indicates that though there are differences in the number of nodes participating in each layer as well in the connections {\it i.e.,} the average connectivity is conserved across all life stages.
Further, the network diameter indicates how much far are the two most distant nodes in a network.
The diameter being small for all the layers suggests that all the nodes are in proximity and the graph is compact \cite{Lewis}. 
The diameter, and thus the compactness, of a PPI network, can be interpreted as the overall easiness of the proteins to communicate or influence their reciprocal function.

\begin{figure}[h]
\begin{center}
\includegraphics[width=12cm,height=7cm]{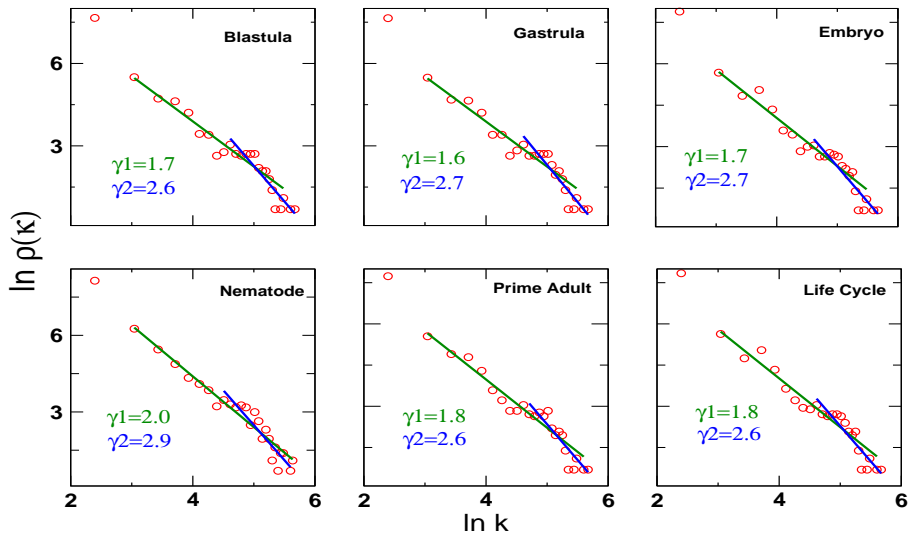}
\caption{(Color On-line) {Degree distribution of different layers of multilayer PPI network of {\it C. elegans} life stages.}  The circles (red) represent the data points and the lines (green and blue) mark power law fit. “$\gamma$” refers to the exponent of power law.}
\label{fig2}
\end{center}
\end{figure}

Further, an intriguing observation of degree distribution {is found} which follows two distinct fitting scales in all the layers {\it i.e.,} two power law in each layer (Fig. \ref{fig2}).
Many network studies have reported an absence of the perfect power law for the overall range of the degree. 
Various real systems show power law in the central part of data only and deviation from it in the small or the large scale. 
Furthermore, the value of first power law exponent $\gamma$ is lower than the second one in all the layers. 
Several models have been used to explain the origin of two power laws found in many systems.
The models include, the geometric Brownian motion model, the preferential attachment model and the generalized model of the creation of new links between old nodes which increases with evolution time \cite{doublePowerlaw}.
These models suggest that the evolution of a network is characterized by two parts {\it i.e.,} (i) a leading ingredient of a network and (ii) fluctuations within existing connections between nodes, being one of the reasons to lead the double power law nature of degree distribution.
Here, the power law nature of PPI layer implicates that robustness of a network is maintained not only by acquisition of new interactions by hub proteins but also by contribution of new or altered interactions within existing proteins for ease of pathway processes \cite{Lorimer} which might have arisen due to the presence of internal physiological and external environmental factors during the development of an organism \cite{EnvFactor}.

What follows that the individual network exhibits overall similar statistics for widely investigated structural properties {\it i.e.,} smaller diameter, and larger average clustering coefficient than the corresponding random networks as well as existence of two power law but the crucial differences among them, are revealed through the analysis of the degree-degree correlation and spectral properties.
All the network layers show overall positive degree-degree correlation (Table \ref{table1}).
{Though most of the biological networks exhibit disassortative nature \cite{Disso}, }a positive degree-degree correlation is observed in many other biological networks \cite{AssortativityBio}.
It is reported that assortative networks are strongly clustered and can have functional modules \cite{Sah}.
A high value of $r$ here suggests the presence of functional modules as functional areas of PPI network, but there is lack of clear evidence to prove whether a functional area forms a functional module \cite{Sah}.
The corresponding random networks have $r$ value close to zero (Table \ref{table2}). 
This is not surprising as the networks with the same average degree and size may still differ significantly in various network features since the nodes are randomly connected and the value of assortativity coefficient of a network is determined by degrees of interacting nodes \cite{PiraveenanM}.
The assortativity observed in PPI networks implicates that overall interaction patterns of nodes having similar degree is conserved  in all the layers.
Next, Blastula and Gastrula PPI layer have same $r$ value which is not surprising as a large number of proteins and, hence their interactions are common in these two layers.
Further, though all PPI layers show $r$ value close to each other, the $r$ value of Nematode is lower than other PPI layers as well as the difference in the $r$ value from that of other layers is very high.
It suggests that Nematode PPI layer is not as assortative as other layers which implicates in the presence of less structural modules in Nematode as of other layers.
Further, $r$ values close to zero of the random networks suggest that the network with random interactions tend to have less $r$ values.
All these suggest that Nematode PPI layer is more random than the other layers.

\begin{table}
\begin{center}
\caption{
{The properties of corresponding ER random networks of each layer of multilayer PPI network.}
Here $\langle C \rangle$, $D$, $r$, $\lambda_{0}$, $O \beta_L$, and $S$ represent average clustering coefficient, diameter, degree-degree correlation coefficient, count of zero eigenvalue, overlap-link betweeness correlation coefficient, von Neumann entropy coefficient for corresponding random networks of each layer, respectively. All values are averaged over 20 realizations of corresponding random networks. $logN$ represents maximum entropy of the network with $N$ size.}
\begin{tabular} {c c c c c c c | c}    
\hline  \specialrule{1pt}{0pt}{0pt}
Network  	     & $\langle C \rangle$    &$D$  & $r$    &$\lambda_{0}$  &$O \beta_L$       &$S$                        &$log N$\\ \hline   \specialrule{1pt}{0pt}{0pt}
Blastula		     & 0.005                           	&5   & 0       &0     &-0.1                          	&11.43 $\mp$0.02 &11.49\\ 
Gastrula		     & 0.005                 	   		&5   & 0       &0     &-0.1	               	   		   &11.39 $\mp$0.04 &11.48\\  
Embryo		     & 0.005                        		&4   & 0       &0      &-0.1	            		       &11.39 $\mp$0.04 &11.80\\
Nematode		 & 0.006                        		&5   & 0       &0      &-0.11	            			&12.12 $\mp$0.04 &12.21\\
Prime Adult	 & 0.005                      		&5   & 0       &0      &-0.09	$\mp$0.04     &11.34 $\mp$0.04 &11.60\\ 
Life Cycle	     & 0.006                            &5   & 0        &0      & -0.13 	          		 	&11.81 $\mp$0.04 &11.74\\ \hline  \specialrule{1pt}{0pt}{0pt}
\end{tabular}
\label{table2}
\end{center}
\end{table}

To get deeper insights into the organization of connections in PPI networks, these networks {are further analyzed} with the weak ties hypothesis.
Here, the links having low overlap in their end nodes are termed as the weak ties and links having high link betweeness centrality are the ones known to be stronger as they help in connecting in different modules \cite{Onnela}.
All the network layers exhibit the negative value of $O\beta_L$ correlation coefficient (Table \ref{table1}) which suggests the presence of weak ties in PPI network.
It is suggested that the complete architecture of PPI network is composed of different biological pathways and metabolic cycles, and  a protein involved in particular pathway plays role in regulating other pathways as well, termed as cross talk between pathways \cite{Yamada}.
The corresponding random networks exhibit negative $O\beta_L$ value (Table \ref{table2}), but PPI layers have more negative $O\beta_L$ value than the corresponding random networks.
It implicates that the network with random interactions tends to have more $O\beta_L$ value.
Therefore, the highest $O\beta_L$ value in Nematode suggests that Nematode PPI layer is more random than the other layers.
Nevertheless, understanding of the evolution of embryogenesis in {\it C. elegans} is still fragmentary, more randomness shown by the less value of assortativity coefficient, and high value of $O\beta_L$ coefficient than other layers may suggest that the presence of more randomness in this layer.
It is reported that the processes of cellular diversification plays vital role in the larval Nematode development \cite{Schulze}, which may be the reason for more randomness in this network layer.


\subsection*{Spectral properties}

So far {the study is} focused on various structural aspects of all the layers which have demonstrated distinguishable structural features of the Nematode from other layers.
Further, the spectra of these networks {are analyzed} since spectra is known to be fingerprint of corresponding network \cite{BookSpectra}.
The network spectra not only provide insight to functional modules \cite{Ott} and randomness in the connection architecture \cite{Rai}, but also relate with the dynamical behavior of the system as a whole \cite{Nishikawa}.
{It is observed that} the spectra of all the PPI layers have high degeneracy at the zero eigenvalue.
The occurrence of degeneracy at zero eigenvalue for PPI layers is not surprising here as many biological and technological networks are known to exhibit high degeneracy at zero eigenvalue \cite{SJalan}.
Interestingly, the number of zero eigenvalue has direct relation with complete and partial duplication of nodes \cite{Alok} as discussed in Material and methods section.
The number of complete duplicates $D_0$ which contributes to the same number of zero eigenvalue ($\lambda_0$), is listed in Table \ref{table1}.
The appearance of duplicate nodes in biological networks have been emphasized to be arising due to the gene duplication process as a consequence of 
evolution \cite{Jost}.
The corresponding random networks do not exhibit degeneracy at zero eigenvalue, indicating that this count of duplicate nodes in PPI layers reside not only in the sheer number of proteins and interactions taking part in particular layer, but also in how individual PPI layer is evolved or designed to fulfill the cellular functions.

An important observation is that despite overall similar spectral properties {\it i.e.} degeneracy at zero eigenvalue as well as shape of eigenvalue distribution \cite{SM}, the height of the peak at zero eigenvalue differs in all the layers.
Since size of the networks differ at different life stages, in order to take care of the impact of size on the occurrence of the zero degeneracy, {the count of zero eigenvalues are normalized} by dividing with $N$ and find that these normalized values also differ in PPI layers (Table \ref{table1}).
What is important here is that the genome of an organism remains same in all the life stages, still there is occurrence of different count of duplicate nodes in PPI layers.
In PPI network, gene duplication is understood as duplication of protein since the duplicated protein is the expressed product of the duplicated gene as well as it is the identical copy of the parent one.
The duplicated protein initially shares common function as of the parent protein which results in the same interaction partners, later it functionally diversifies to acquire different interaction partners \cite{Ispolatov}.
Since every protein contributes to the specific physiological and developmental process, and different physiological and developmental processes in each life stage would require different set of proteins, may result in different count of duplicate nodes.
Taken together, it suggests that there may be the role of specific biological responses at each developmental stage in the process of gene duplication.
Further, Nematode PPI layer exhibits more number of zero eigenvalues than other layers.
It is reported that Nematode is a crucial stage for cellular diversification and organogenesis, as well as there are developmental transitions during continuous and interrupted larval Nematode development \cite{Rougvie}, which might result in more duplicates in Nematode than other life stages.

Furthermore, the von Neumann entropy of all the layers {are calculated and observed that there are} different $S$ values in each PPI layer (Table \ref{table1}).
To get more insights on this, entropy $S$ of each PPI layer {is compared} with the maximum entropy $logN$ of that layer, also with $S$  of the corresponding random network.
Firstly, PPI layers as well as random networks display lesser entropy than corresponding maximum entropy which is quite intuitive, since any network with $N$ size can have maximum $logN$ entropy.
Secondly, corresponding random networks display higher entropy than PPI layers.
The networks with random interactions among nodes tend to show higher entropy.
The comparison of entropy of each PPI layer with the maximum entropy as well as with the corresponding random network indicates presence of varying complexity in each PPI layer.
It recites the similar notion of varying complexity present in each layer which is deduced earlier by structural features.
It is potentially important since it may be in consequence with the specific developmental and evolutionary stimuli associated with each life stage.
Further, Nematode has the highest $S$ value which suggests the presence of more complexity in this layer than all other PPI layers.
Taken together, more complexity in Nematode than all other PPI layers and the least $r$ value and the highest $O\beta_L$ suggest in more randomness in this PPI layer.
As {it is} discussed earlier, the contribution of specific developmental factors at larval Nematode development may result here in more complexity in PPI layer of Nematode life stage than all other PPI layers.

\section*{Conclusions}

The proteome analysis of each layer of {\it C. elegans} multilayer PPI network exhibits the overall similarity in structural features such as smaller diameter and larger average clustering coefficient than the corresponding random networks as also observed for other biological networks  \cite{BarabasiReview, Clauset}.
The degree distribution following power law in each of the PPI layer is indicative of the robustness {of the underlying system}.
Although the widely studied structural properties exhibit similar statistics, {the} crucial differences in the network layers through the analysis of the degree-degree correlation and spectral properties, which also turn out to be of potential importance in understanding varying complexity in each layer.
The values of $r$, $O\beta_L$ and $S$ coefficients of each layer ubiquitously behave in similar {manner} and are found to be comparable with other PPI layers of underlying system.
Interestingly, the layer of Nematode life stage exhibits notable distinguishing properties than other layers, which in overall indicates that this layer being the most random among all the layers.
Further, each PPI layer exhibits different degeneracy at the zero eigenvalue which is related to node duplication, suggesting the role of specific biological responses at each developmental stage in the process of gene duplication.

To summarize, an extent of varying complexity is observed in the organization of PPI networks of individual layers of multilayer PPI network.
It recites the fact that biological complexity arises at several levels in the development of {\it C. elegans} such as, from single cell embryo to multicellular completely developed organism where each life stage is associated with different physiological and molecular changes \cite{Zoltan}.
The varying complexity observed in life stages can further be used to understand and capture important developmental changes in an organism.

\pagebreak

\section*{Acknowledgment}
SJ is grateful to Department of Science and Technology (DST), Government of India and Council of Scientific and Industrial Research (CSIR), Government of India project grants EMR/2014/000368 and 25(0205)/12/EMR-II for financial support, respectively. PS acknowledges DST for the INSPIRE fellowship (IF150200) as well as the Complex Systems Lab members for timely help and useful discussions.

\newpage

\cleardoublepage

\end{document}